\documentclass[12pt]{article}
\usepackage{graphicx}
\usepackage{psfig}
\usepackage{epsf}
\usepackage{amssymb}
\newcommand{\ba}{\begin{array}}
\newcommand{\ea}{\end{array}}
\newcommand{\bd}{\begin{displaymath}}
\newcommand{\ed}{\end{displaymath}}
\newcommand{\be}{\begin{equation}}
\newcommand{\ee}{\end{equation}}
\newcommand{\bea}{\begin{eqnarray}}
\newcommand{\eea}{\end{eqnarray}}
\newcommand{\Rsl}{{\not \! \!{R}}}

\def\n{\nu}

\def\etal {{\em et al.,\ }}
\def\th13 {\theta_{13}}

\begin{document}
\thispagestyle{empty}
\begin{flushright}
\texttt{hep-ph/0608034}\\
\texttt{HRI-P-06-08-001}\\
\texttt{CU-PHYSICS-14/2006}\\
\end{flushright}
\vskip 15pt

\begin{center}
{\Large {\bf  Can R-parity violating  supersymmetry 
be seen in long baseline beta-beam
experiments?}}

\vskip 15pt
\renewcommand{\thefootnote}{\alph{footnote}}

\large{\bf{
Rathin Adhikari $\footnote{E-mail address: 
rathin\_adhikari@yahoo.com}^{\dagger}$, 
 Sanjib Kumar Agarwalla$\footnote{
E-mail address: sanjib@mri.ernet.in}^{\ddagger}$ 
}}\\ 
\large{\bf{ and Amitava
Raychaudhuri$^{\ddagger}$}}
\end{center}
\begin{center}
\small$\phantom{i}^{\dagger}${\em Jagadis Bose National Science
Talent Search,\\ 1300 Rajdanga Main Road, Kasba, \\
Kolkata - 700107, India }\\ 
\vskip 10pt
\small$\phantom{i}^{\ddagger}${\em Harish-Chandra Research Institute,\\ 
Chhatnag Road, Jhunsi, Allahabad - 211019, India \\ 
and\\
Department of Physics,
University of Calcutta, \\ Kolkata - 700009, India}
\vskip 30pt

{\bf ABSTRACT}

\vskip 0.5cm

\end{center}

Long baseline oscillation experiments may well emerge as test
beds for neutrino interactions as are present in R-parity
violating supersymmetry.  We show that flavour diagonal (FDNC)
and flavour changing (FCNC) neutral currents arising therefrom
prominently impact  a neutrino $\beta$-beam experiment with the
source at CERN and the detector at the proposed India-based
Neutrino Observatory.  These interactions may preclude any
improvement of the present limit on $\theta_{13}$ and cloud the
hierarchy determination  unless the upper bounds on ${\not
\!\!R}$ couplings, particularly $\lambda^{\prime}$, become
significantly tighter. If ${\not\! \!R}$ interactions are independently
established then from the event rate a lower bound on
$\theta_{13}$ may be set. We show that there is scope to see a
clear signal of non-standard FCNC and FDNC interactions,
particularly in the inverted hierarchy scenario and also
sometimes for the normal hierarchy.  In favourable cases, it may
be possible to set lower and upper bounds on $\lambda^{\prime}$
couplings.   FCNC and FDNC interactions due to $\lambda$ type
${\not\! \!R}$ couplings are unimportant. 


\newpage

\renewcommand{\thesection}{\Roman{section}}
\setcounter{footnote}{0}
\renewcommand{\thefootnote}{\arabic{footnote}}

\section{Introduction}
Neutrino physics, a bit player on the physics stage in
yesteryears, has now donned a central role. Various experiments
on atmospheric \cite{atm}, solar \cite{solar}, reactor
\cite{reac}, and long baseline neutrinos \cite{base} not only
indicate oscillation but also pin down most neutrino mass and
mixing parameters -- two mass-square differences and two mixing
angles.  The best-fit \cite{bestfit} values with $3\sigma$ error
from atmospheric neutrinos are $|\Delta m^{2}_{23}|\simeq
2.12^{+1.09}_{-0.81}\times 10^{-3}$ eV$^2$, $\theta_{23}\simeq$
${45.0^\circ}^{+10.55^\circ}_{-9.33^\circ}$ and  from solar
neutrinos  $\Delta m^{2}_{12} \simeq 7.9^{+1.0}_{-0.8} \times
10^{-5}$ eV$^2$, $\theta_{12}\simeq$ ${33.21^\circ
}^{+4.85^\circ}_{-4.55^\circ}$. Here   $\Delta m^{2}_{ij}$=
$m^{2}_{j} - m^{2}_{i}$. The  sign of $\Delta m^{2}_{23}$ is yet
unknown and the neutrino mass spectrum will be referred here
as normal (inverted) hierarchical if it is positive (negative).
Using reactor antineutrinos \cite{bestfit,reac,theta13},  an upper
bound has been set on the third mixing angle at $3\sigma$  level
as $\sin^2\theta_{13}   < 0.044$   resulting in $\theta_{13} <
12.1^\circ$.  The phase $\delta$ in the neutrino mixing matrix is
not known.

Research in this area is now poised to move into the precision
regime. Can we use upcoming neutrino experiments to probe
non-standard interactions like $\Rsl$ supersymmetry? If present,
can they become spoilers in attempts to further sharpen the
neutrino properties?  We attempt to address these issues with a
specific experiment as our laboratory.

Long baseline neutrino oscillation experiments using neutrino
factories    \cite{md} and  $\beta$-beams
\cite{bb1, bb2, bb3, bb4} hold promise of refining our
knowledge of $\theta_{13},  \; \delta,$ and the sign of
$\Delta m_{23}^2$.  A possible experiment in this category would
use neutrinos from a  $\beta$-beam from CERN to the proposed
India-based Neutrino Observatory \cite{ino} (INO), a baseline  of
$\sim$ 7152 Km. This is the set-up which we
consider here. For such an experiment the $\beta$-beam is
required to be boosted to high $\gamma$.

Interaction of neutrinos with matter affect 
long baseline experiments and this becomes more prominent
at higher values of $\theta_{13}$.
Various    authors \cite{matter} have considered this effect for
atmospheric neutrinos. 

Apart from the electroweak effects, there may well be
non-standard interactions leading to flavour changing as well as
flavour diagonal neutral currents (FCNC and FDNC).  Here we have
in mind interactions with quarks and leptons involving an initial
and a final neutrino. If there is no change in the neutrino
flavour -- as, for example, in $Z^0$ exchange --  this is
classified as an FDNC process, while it would be FCNC otherwise.
R-parity violating supersymmetric models   (RPVSM)
\cite{rparity}, which have such interactions already
built-in\footnote{{\em e.g.} through  squark ($\lambda^\prime$-type
couplings) or slepton ($\lambda$-type couplings) exchange.},
will be the main focus of our work.   Very recently a model in
which couplings associated with FCNC and FDNC can be quite a bit
higher than permitted in RPVSM has also been considered
\cite{david, fcnc1}. Naturally, here the
matter effect  will be further enhanced. However, as RPVSM is a
well-studied, renormalizable  model which  can satisfy all
phenomenological constraints currently available, we shall
restrict our main analysis only to it and shall make
qualitative remarks about the other model,
for which our results can be easily extended.

Consequences of FCNC and FDNC for solar and atmospheric neutrinos
\cite{solar1, atm1}, and neutrino factory experiments \cite{fac1}
have been looked into.  Our focus is on $\beta $-beam
experiments, particularly over a long distance (7152 Km)
baseline.  Our analysis encompasses both normal and inverted
hierarchies and we also  incorporate all relevant trilinear
R-parity violating couplings leading to FCNC and FDNC. Huber {\em
et al} \cite{huber} have a somewhat similar analysis  using
neutrino beams obtained from muon decays.

The very long baseline from CERN to INO will capture a
significant matter effect and offers a scope to signal
non-standard interactions.  We examine whether the presence of
$\Rsl $ interactions will come in the way of constraining the
mixing angle $\theta_{13}$ or unraveling the neutrino mass
hierarchy.  The possiblity to obtain bounds on some 
R-parity violating couplings is also probed.  

\section{$\beta$-beams}

A beta-beam \cite{bb1}, is a  pure, intense, collimated beam of
electron neutrinos or their antiparticles produced {\em via} the
beta decay of completely ionized, accelerated radioactive ions
circulating in a storage ring \cite{jacques}.  It  has been
proposed to produce $\nu_e$ beams through the decay of highly
accelerated  $^{18}$Ne ions and ${\bar \nu_e}$ from $^6$He
\cite{jacques}. More recently, $^{8}$B and $^{8}$Li \cite{rubbia}
with much larger end-point energy have been suggested as
alternate sources since these ions can yield higher energy
$\nu_e$  and $\bar\nu_e$ respectively, with lower values of  the
Lorentz boost $\gamma$.  It may be possible to have both beams in
the same ring, an arrangement which will result in a $\nu_e$ as
well as a $\bar\nu_e$ beam pointing  towards a distant target.
In such a set-up the ratio between the boost factors of the two
ions is fixed by the $e/m$ ratios of the ions.   Here, we will
present our results with the $^{8}$B ion ($Q = 13.92$ MeV  and
$t_{1/2} = 0.77 s$) taking $\gamma = 350$.  As we show,  $\gamma
\sim 350$ may permit a distinction between matter effects due to
Standard Model interactions and those from R-parity violating
supersymmetry (SUSY). Details of the neutrino flux from such a
$\beta$-beam can be found in \cite{bb4}.

Using the CERN-SPS at its maximum power, it will be possible to
access $\gamma$ $\sim$ 250\footnote{$\gamma$ = 250 yields too few
events in this experiment for the extraction of interesting physics.}
\cite{gamma_250}.  For a medium $\gamma$ $\sim$ 500, a
refurbished SPS with super-conducting magnets or an acceleration
technique utilizing the LHC
\cite{gamma_250,burguet,LHC_upgrade} would be required.  In the
low-$\gamma$ configuration, $1.1\times 10^{18}$ useful decays per
year can be obtained with $^{18}$Ne ions
\cite{autin,terr}.  We have used this same luminosity for $^{8}$B
and higher $\gamma$ \cite{donini}.  This issue is being examined in an on-going
dedicated machine study at CERN.

\section{R-parity violating Supersymmetry}

In supersymmetric theories \cite{rparity}, gauge invariance does
not imply baryon number  (B) and lepton number (L) conservation
and, in general, R-parity (defined as R $= (-1)^{3B + L + 2S}$
where $S$ is the spin) is violated.  To maintain consistency with
non-observation of fast proton decay etc, in the R-parity violating
Minimal Supersymmetric Standard Model (imposing baryon number
conservation) one may consider the following superpotential
\bea
W_{\not L} =  
\sum_{i,j,k} \frac{1}{2}\lambda_{ijk} {L}_i {L}_j {E}_k^c +
\lambda_{ijk}' {L}_i {Q}_j {D}_k^c + \mu_i {L}_i {H}_u ,
\label{super}
\eea
(suppressing colour and $SU(2)$ indices) where $i,\;j,\;k$ are
generation  indices.  Here ${L}_i$ and ${Q}_i$ are $SU(2)$-doublet lepton
and quark superfields respectively; ${E}_i$, ${D}_i$ denote the
right-handed $SU(2)$-singlet charged lepton and down-type quark
superfields respectively; ${H}_u$ is the Higgs superfield which
gives masses to up-type quarks. Particularly, $\lambda_{ijk}$ is
antisymmetric under the interchange of the first two generation
indices. We assume that the bilinear terms have been rotated away
with appropriate redefinition of superfields and focus on the
two trilinear L-violating terms with $\lambda$ and
$\lambda^\prime$ couplings.  Expanding those in standard
four-component Dirac notation, the  quark-neutrino interaction
lagrangian  can be written as:
\bea
{\cal {L_{\lambda'}}} =  \lambda'_{ijk} ~\big[ ~\tilde d^j _L \,\bar d ^k _R \nu^i _L
  + (\tilde d ^k_R)^\ast ( \bar \nu ^i_L)^c d^j _L   \big] +
  h.c.
\label{lag_1}
\eea
Above, the sfermion fields are characterized by the tilde sign. The
charged lepton interacts with the neutrino {\em via}
\bea
{\cal {L_{\lambda}}} &=&  \frac{1}{2}\lambda_{ijk} ~\big[ ~\tilde
e^j _L \,\bar e ^k _R \nu^i _L + (\tilde e ^k_R)^\ast (\bar \nu
^i _L)^c e^j _L - (i \leftrightarrow j) \big] + h.c,
\label{lag_2}
\eea
In what follows, we
shall consider these couplings as real but will entertain both positive
and negative values.   The interactions of neutrinos with electrons and 
$d$-quarks  in matter induce transitions    
(i) $\n_i + d \rightarrow \n_j + d $ 
and (ii)
 $\n_i + e \rightarrow \n_j
+e $. (i) is possible through $\lambda^{\prime}$ couplings {\em via} 
squark exchange  for all $i, j$
and through $Z$ exchange for $ i = j$ 
while (ii)  can proceed  {\em via} $W$ and $Z$ exchange for $i = j$,
as well as    through $\lambda$ couplings  {\em via} slepton
exchange for all $i, j$. 

\section{Neutrino oscillation probabilities \&  matter effect}

 In a three neutrino framework, the neutrino flavour states $|\nu_\alpha
\rangle$, $\alpha = e, \mu, \tau$, are related to
the neutrino mass eigenstates $|\nu_i\rangle$, $i=1,2,3$, with
masses $m_i$:
\begin{equation}
\vert\nu_\alpha \rangle = \sum_i U_{\alpha i} \vert\nu_i\rangle~,
\end{equation}
where $U$ is a  $3 \times 3$ unitary matrix which can be
expressed as:
\be
U = V_{23} V_{13} V_{12} ,
\ee
where
\bea
V_{23} = \left(
          \begin{array}{ccc}
          1 & 0 & 0  \\
 0 & c_{23}
&  s_{23}\\
  0 & -  s_{23} & c_{23}
\end{array} \right) ;\;\; 
V_{13} = \left(
          \begin{array}{ccc}
          c_{13} & 0 & s_{13}e^{-i\delta}  \\
 0 & 1
& 0 \\
 s_{13}e^{i\delta} & 0
& c_{13} \end{array} \right)
; \;\; V_{12} = \left(
          \begin{array}{ccc}
          c_{12} & s_{12} & 0  \\
 - s_{12}  & c_{12} 
&  0\\
  0 & 0 & 1
\end{array} \right) .
\label{e:mns}
\eea
$c_{ij}=\cos\theta_{ij}$, $s_{ij}=\sin\theta_{ij}$ and
$\delta$
denotes the $CP$ violating (Dirac) phase. There may be a diagonal
phase matrix on the right containing two more Majorana phases.
These are not considered below. Apart from
some qualitative remarks, we  present our result considering
$\delta = 0$ corresponding to the $CP$ conserving case.  In the
mass basis of neutrinos

\be 
     M^2 = {\rm{diag}}(m_1^2,m_2^2,m_3^2) = U^\dagger M_\nu^+ M_\nu U,
\label{e:m1}
\ee 
where $M_\nu$ is the neutrino mass matrix in the flavour basis and $m_1, 
m_2$, and  
$m_3$ correspond to masses of three neutrinos in the ascending
order of magnitudes respectively for the normal hierarchy. If
$m_2 > m_3$, we have an  inverted hierarchy.

The neutrino flavour eigenstates evolve in time as: 

\bea
i\frac{d}{d t}\left(
\begin{array}{c}
\n_e(t)\\
\n_{\mu}(t)\\
\n_{\tau}(t)  \end{array} \right)  
=  
H \left(
\begin{array}{c}
\n_e(t)\\
\n_{\mu}(t)\\
\n_{\tau}(t)  \end{array} \right),
\eea
where 
\be
H = E \times {\bf 1}_{3 \times 3} + 
U \left( { M^2 \over 2 E} \right) 
   U^\dagger 
+  R .
\label{e:m2}
\ee
Here $E$ is the neutrino energy while ${\bf 1}_{3 \times 3}$ is the 
identity matrix.  
$R$ is  a $3 \times 3$ matrix reflecting the matter effect,
absent for propagation in vacuum. 
\be
R_{ij} = R_{ij}(SM) + R_{ij}(\lambda^{\prime}) + R_{ij}(\lambda).
\ee
Specifically, 
\be
R_{ij}(SM)   = \sqrt{2}  G_F n_e \delta_{ij} (i,j=1)+ {G_F n_n \over \sqrt{2}} 
 \delta_{ij} ,
\label{e:rsm}
\ee
\be
R_{ij}(\lambda^{\prime})  =  \sum_m \left( {\lambda_{im1}^\prime \lambda_{jm1}^{\prime} 
\over 4 m^2 ({\tilde d_m})} n_d + {\lambda_{i1m}^{\prime  } \lambda_{j1m}^{\prime} 
\over 4 m^2({\tilde d_m})  } n_d \right) ,
\label{e:rlp}
\ee
\be 
R_{ij}( \lambda ) = 
\sum_{k \neq i, j}  {    \lambda_{ik1} \lambda_{jk1} \over
 4 m^2 ({\tilde { l^{\pm}_k}}) }  n_e
 + \sum_n  {    \lambda_{i1n} \lambda_{j1n} \over
 4 m^2 ({\tilde { l^{\pm}_n}}) }  n_e ,
\label{e:rl}
 \ee  
where $G_F$ is the Fermi constant, $n_e$, $n_n$, and $n_d$,  
respectively, are the electron, neutron,  and down-quark densities in
earth matter. Note that $R$ is a symmetric matrix and also that
antineutrinos will have an overall opposite sign for $R_{ij}$.
Assuming  earth matter to be isoscalar, $ n_e = n_p =n_n $ and  $
n_d = 3 n_e $.  The current bounds on the $\Rsl$ couplings
\cite{rparity} imply that the $\lambda^\prime$ induced contributions
to $R_{11}$, $R_{12}$ and $R_{13}$ are several orders less than
$\sqrt{2} G_F n_e$. We neglect those terms in our analysis.  The
upper bounds on all  couplings in $R_{ij}(\lambda)$ are also very
tight \cite{rparity} in comparison to $\sqrt{2} G_F n_e$ and
their effect will be discussed later.
So, first we consider, in addition
to the Standard Model contribution, only
\bea
R_{23} &=& R_{32}= {n_d \over 4 m^2 ( {\tilde d_m} )}
 \left( \lambda_{2m1}^{\prime }
  \lambda_{3m1}^{\prime} +
  \lambda_{21m}^{\prime}
  \lambda_{31m}^{\prime} \right) ,  \nonumber \\
R_{22} &=& {n_d \over 4 m^2 ( {\tilde d_m} )}
 \left( \lambda_{2m1}^{\prime \; 2}
 + \lambda_{21m}^{\prime \; 2} \right)   \; ,\;\; 
R_{33} = {n_d \over 4 m^2 ( {\tilde d_m})}
 \left( \lambda_{3m1}^{\prime \; 2}
 + \lambda_{31m}^{\prime \; 2} \right) ,  
\label{e:R1}
\eea
which are comparable to $\sqrt{2} G_F n_e$.  One  can see from
eq. (\ref{e:R1}) that $R_{23} \neq 0$  implies both $R_{22}$ and
$R_{33}$ are non-zero\footnote{However,
in other models \cite{david} this may not be the case.}.

The current bounds on the relevant couplings are as follows \cite{rparity}:
\be
|\lambda_{221}^{\prime}, \lambda_{231}^{\prime}|  < 0.18 ;  \;
|\lambda_{21m}^{\prime}|  < 0.06 ; \; |\lambda_{331}^{\prime}|  <
0.58 ; \; |\lambda_{321}^{\prime}|  <  0.52 ; \;
|\lambda_{31m}^{\prime}| < 0.12 ,
\label{e:lim}
\ee
for down squark mass  $m_{\tilde d} = 100$ GeV.  The chosen
limits on $\lambda_{21m}^{\prime}$ and $\lambda_{31m}^{\prime}$
do not conflict with the ratio $R_{\tau
\pi}=\Gamma (\tau \rightarrow \pi \nu_{\tau})/\Gamma  (\pi
\rightarrow \mu \bar{\nu}_{\mu} ) $ \cite{rparity}. However, the
recently published BELLE bound  on the mode $\tau
\rightarrow \mu \pi^0$ \cite{belltau} tightly constrains
precisely those products of the $\lambda^\prime$ couplings which
enter in $R_{23} = R_{32}$ in eq.
(\ref{e:R1}). It has been shown that $|{\lambda'_{21m}
\lambda'^{\ast} _{31m}}|$ and  $|{\lambda'_{2m1}
\lambda'^{\ast} _{3m1}}|$ both must be $< 1.8 \times 10^{-3}
(\frac{\tilde m}{100~{\rm GeV}})^2$ \cite{rpartau}. This 
effectively makes $R_{23}$ negligible
for our purposes.

In general, it is cumbersome to write an analytical form of the
probability of neutrino oscillation in the three-flavour scenario
with matter effects.  However, under certain reasonable
approximations it is somewhat tractable.  Firstly, a constant
matrix can be extracted from $M^2$ in eqs. (\ref{e:m1}) and
(\ref{e:m2}). Also for the energies and baselines under
consideration, $\Delta m_{12}^2 L/E << 1 $.   
Under this approximation, the $V_{12}$ part of $U$ drops out from
eq. (\ref{e:m2}).  With these modifications, the effective mass squared
matrix is:
\be
{{\tilde M}^2 \over 2E}
 = H - (E + {G_F n_n \over \sqrt{2}}) {\bf 1}_{3 \times 3}.
\ee
In the special case
where $R_{22} = R_{33}$, if one uses the best-fit value of the
vacuum mixing angle $\theta_{23} = \pi/4$ then the neutrino mass
squared eigenvalues are:
\be
\left( {\tilde M_2}^2 \over 2E\right)  = R_{22} - R_{23}, \;\; 
\left( {\tilde M_{1,3}}^2 \over 2E\right)  = {1 \over 2} 
\left(  {\Delta m_{13}^2 \over 2E} +R_{11} + R_{22} + R_{23} \mp
A \right) ,
\label{e:Msq}
\ee
where
\be A = {\left[ {\left( {\Delta m_{13}^2 \over 2E} \right) }^2
+ {\left(  -R_{11} + R_{22} + R_{23} \right)}^2  - 2 {\Delta
m_{13}^2 \over 2E}  \cos 2 \theta_{13}\left( R_{11} - R_{22}
-R_{23} \right)  \right]}^{1/2} .
\ee
The matter induced neutrino mixing matrix is given by
\be
U^m = \left( 
\begin{array}{ccc}
U_{11}^m & 0 & U_{13}^m\\
N_1 & -{1 \over \sqrt{2}} & N_3 \\
N_1 & {1 \over \sqrt{2}} & N_3 
\end{array}
\right).
\label{e:U}
\ee
Here
\be
N_{1,3} =  {\left[  {2  {\left ( 
-R_{11} + R_{22} + R_{23} +  {\Delta m_{13}^2 \over 2E} \cos 2\theta_{13}
\pm A \right)}^2 
\over {\left( {\Delta m_{13}^2 \over 2E} \right) }^2 \sin^2 2 \theta_{13} } +
2 \right]}^{-1/2} ,
\ee
where  $N_1$ ($N_3$) corresponds  to +(--) sign in the above expression.
Neglecting the $CP$ phase in
the standard parametrization of $U^m$, one may write $U_{13}^m =
\sin \theta_{13}^m$ and $U_{23}^m =
\sin \theta_{23}^m \cos \theta_{13}^m$. From eq. (\ref{e:U}) it follows
that $\theta_{23}^m = \theta_{23}$, the vacuum mixing angle.
$\theta_{13}^m$, on the other hand, changes from its vacuum value
and it is $\pi/4$ for
\be
R_{11} - R_{22} -R_{23} = {\Delta m_{13}^2 \over 2E} \cos 2 \theta_{13}.
\label{e:mmax}
\ee
 In the absence of non-standard interactions, $R_{22} = R_{23}
=R_{33} =0$ and $R_{11} = \sqrt{2} G_F n_e$,  this is the well-known
condition for matter induced maximal mixing.  Since in eq. (\ref{e:U})
$U_{12}^m = 0$, in the 
$\n_e $ to $\n_{\mu}$ oscillation probability the terms involving 
$({\tilde M_2}^2 - {\tilde M_1}^2)$ 
and $({\tilde M_3}^2 - {\tilde M_2}^2)$ 
will not survive and we get:
\be
P_{\n_e \rightarrow \n_{\mu}}  = 4 \;\;{\left(U_{13}^m\right)}^2\;\; 
{\left(U_{23}^m\right)}^2 \;\;\sin^2 \left(  1.27 \;A \;L \right),
\ee
where $E$, $\Delta m_{13}^2$ and $L$ are expressed in GeV, eV$^2$, and 
Km, respectively. This expression is also valid
for  antineutrinos.  
Using eqs. (\ref{e:Msq}) and  (\ref{e:U}) one can easily obtain
the oscillation probabilities for other channels.

\begin{figure}[htb]
\hskip -0.4cm 
\hskip 8.63cm
\psfig{figure=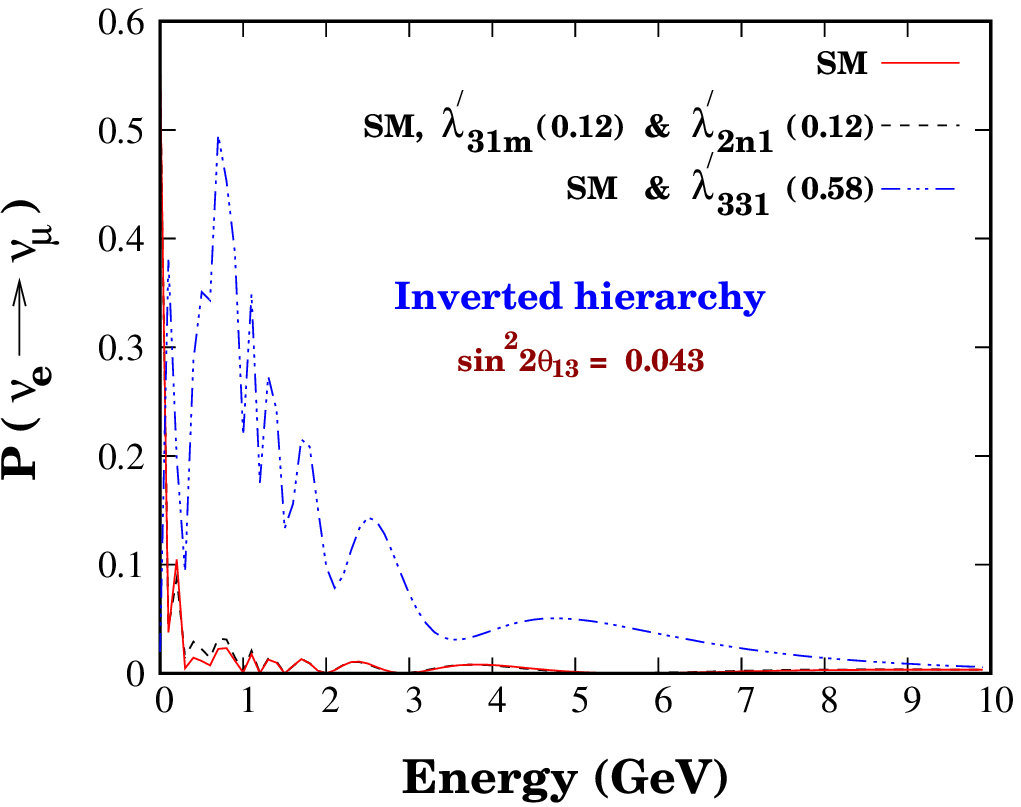,width=8.0cm,height=7.0cm,angle=0}
\vskip -7.0cm
\psfig{figure=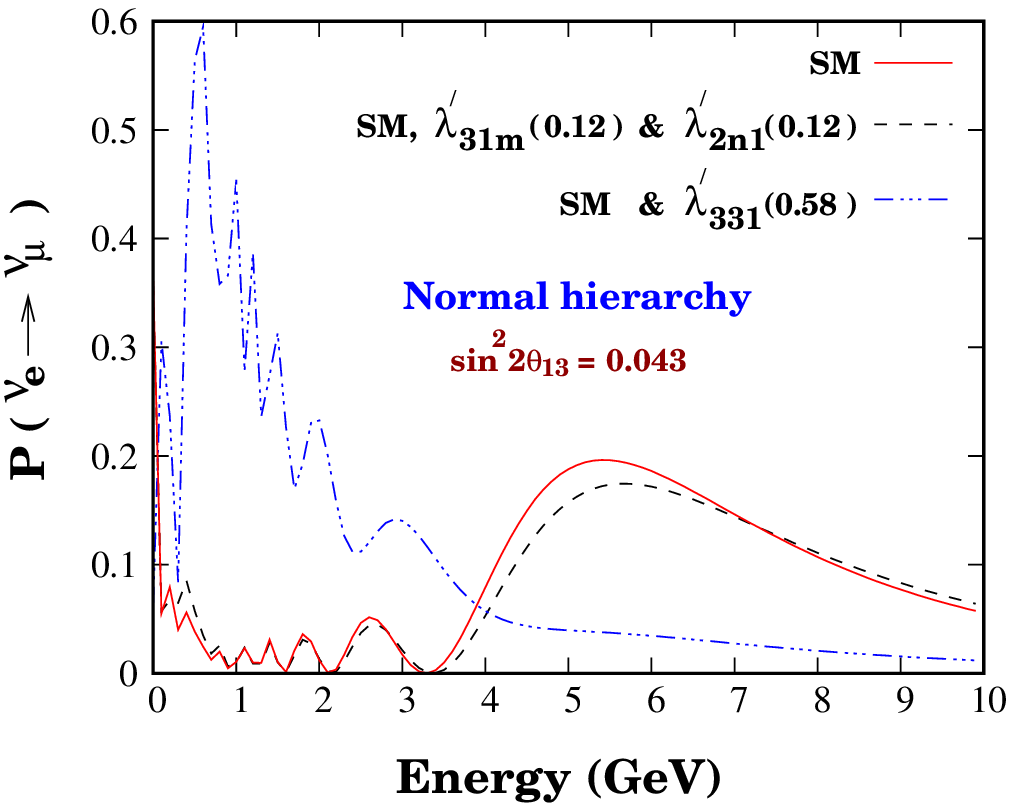,width=8.0cm,height=7.0cm,angle=0}
\caption{ \sf \small
$P_{\n_e \rightarrow \n_{\mu} }$ 
for the normal and inverted mass hierarchies. SM corresponds
to only standard electroweak interactions. The values of
$\lambda^{\prime}$ are given in parantheses. $m$ can take any
value, $n =$ 2 or 3.}
\label{f:osc}
\end{figure}

We use the above analytical formulation as a cross-check on our
numerical results. For example, Fig. \ref{f:osc}, which shows the variation
of $P_{\n_e \rightarrow \n_{\mu} }$ as a function of the energy,
is obtained using the full matter-induced three-flavour neutrino
propagation including non-standard interactions. The
range of energy is chosen in line with the discussions in the
rest of the paper. The probability falls with decreasing $\theta_{13}$ and,
for illustration, we have chosen a value in the middle of its
permitted range. The purpose of Fig. \ref{f:osc} is twofold: (a) to show
how the distinguishability between the normal and  inverted
hierarchies may get blurred by the RPVSM interactions, and (b) how
irrespective of the hierarchy chosen by Nature the results may be
completely altered by the presence of these interactions. Each
panel of Fig. \ref{f:osc} has three curves: the solid line (only electroweak
interactions), dot-dashed line (in addition, $R_{33}$ gets a
non-zero RPVSM contribution), and dashed line 
($R_{22} = R_{33}$  are nonzero, in addition to the
electroweak contribution). Only in the
last case is the analytical formula we have presented above
applicable. We find excellent agreement. Two aspects of the results
are worth pointing out.

First, in the absence of non-Standard interactions, for an
inverted hierarchy the resonance condition eq. (\ref{e:mmax}) is not
satisfied and the oscillation probability is negligible (right
panel solid line).  This could be altered prominently by the RPVSM
interactions (dot-dashed curve) so that the distinguishability
between the two hierarchy scenarios may well get marred by $\Rsl$
SUSY.

Secondly, for the normal hierarchy, it is seen that the peak in
the probability may shift to a different energy in the presence of the
RPVSM interactions. This is because the condition for maximal
mixing in eq. (\ref{e:mmax}) is affected by the $\Rsl$
interactions.  For the inverted hierarchy, the oscillation
probability is considerably enhanced for some energies.  Thus,
physics expectations for both hierarchies will get affected by
RPVSM.

In the following section we dwell on the full impact of this
physics on a long baseline $\beta$-beam experiment.

\section{Results}

We consider a long baseline experiment with a $\nu_e$ $\beta$-beam
source. $\beta$-beams producing $\bar{\nu}_e$ are also very
much under consideration. Broadly speaking, the results obtained
for a $\nu_e$ beam with a normal (inverted) hierarchy are similar
to that with a $\bar{\nu}_e$ beam for an inverted (normal)
hierarchy but details do differ.

The average energy of the $\nu_e$ beam in the lab frame is
$\langle E \rangle = 2 \gamma E_{cms}$, where $E_{cms}$ is the
mean center-of-mass neutrino energy. With $\gamma = 350 $  and
$Q$ = 13.92 MeV for an $^8$B source, $\langle E
\rangle \simeq$ 5 GeV. The proposed ICAL detector at INO
\cite{ino} consists of magnetized iron slabs with
glass resistive plate chambers as interleaved active detector
elements. We present results for a 50 Kt iron detector with
energy threshold 2 GeV. As signature of $\n_e \rightarrow
\n_{\mu}$ oscillation, prompt muons will appear\footnote{The
$\nu_e \rightarrow \nu_\tau
\rightarrow \tau \rightarrow \mu$ route is suppressed by phase space
for $\tau$ production and the branching ratio for the
decay. It contributes at a few per cent level to the signal.}. Their track
reconstruction will give the direction and energy of the
incoming neutrino. ICAL has good  charge identification
efficiency ($\sim 95 \% $) and a good energy resolution $\sim$
10\% above 2 GeV.  Details about the detector    and neutrino
nucleon cross sections may be found in \cite{bb4}.    For the
cross section we include contributions from quasi-elastic, single
pion, and deep inelastic channels.  For our chosen high threshold (2
GeV), the contribution from the deep inelastic channel is
relatively large. For the CERN-INO baseline, the averaged matter
density is 4.21 g cm$^{-3}$.  We use best-fit values of vacuum
neutrino mixing parameters as mentioned in the Introduction. All
the presented results  are based on a five-year ICAL data
sample\footnote{ Backgrounds can be eliminated by imposing
directionality cuts. The detector is assumed to be of perfect
efficiency.}.  

At the production and detection levels, FCNC and
FDNC effects can change the spectrum and detection cross sections
by a small (\raisebox{-.6ex}{\rlap{$\sim$}}\raisebox{.6ex}{$<$}
0.1\%) amount but this would not alter the conclusions. At the
source and detector, they may also mimic the oscillation signal
itself, but these effects are tiny\footnote{This is due to the
very tight constraints from $\mu \rightarrow e$ transition limits
in atoms \cite{mu2e}.} ($\sim {\cal O}(10^{-14})$). Here we
discuss how FCNC and FDNC may significantly modify the
propagation of neutrinos  through matter over large distances.

\subsection{Extraction of $\theta_{13}$ and determination of hierarchy}

\begin{figure}[thb]
\psfig{figure=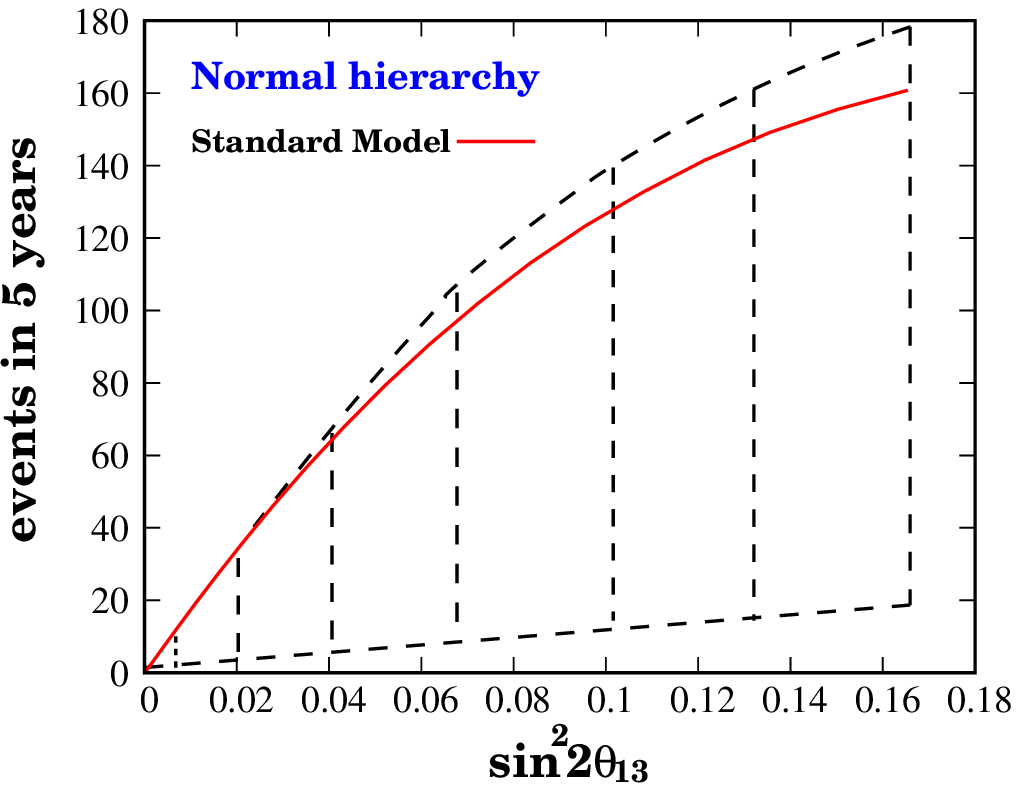,width=8.0cm,height=7.0cm,angle=0}
\vskip -7.00cm
\hskip 8.43cm
\psfig{figure=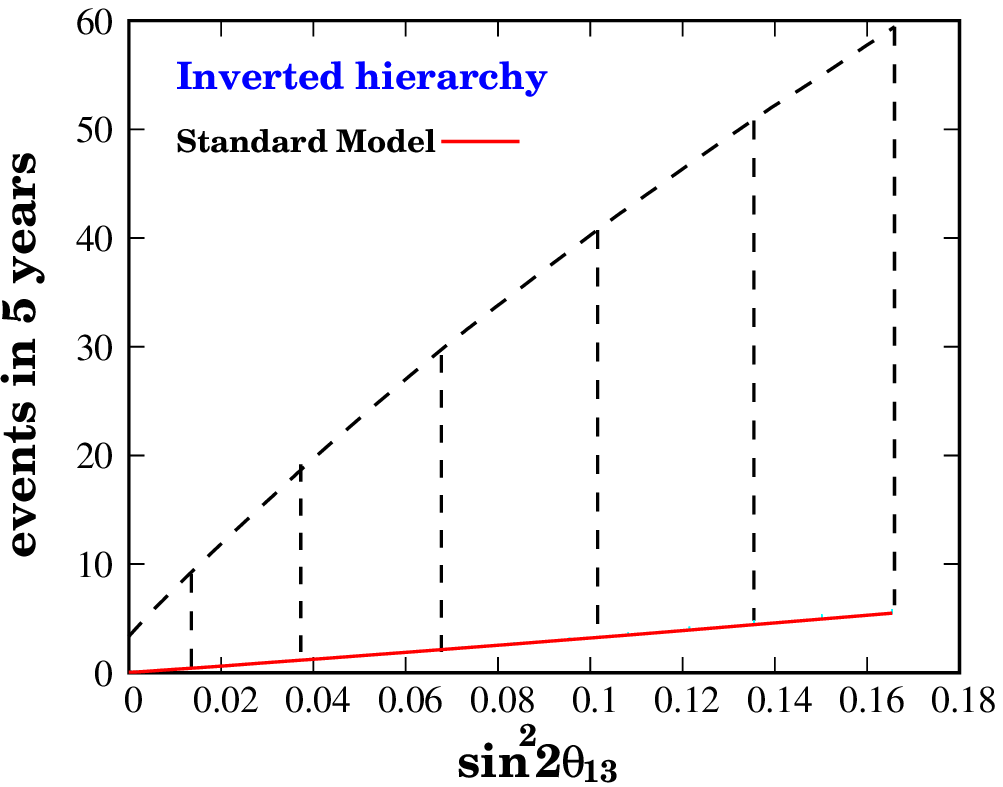,width=8.0cm,height=7.0cm,angle=0}
\caption{\sf \small
Number of muon events for normal and inverted mass hierarchies as
a function of $\sin^2 2\theta_{13}$ for a five-year ICAL run.
The solid lines correspond to the absence of any non-Standard 
neutrino interaction. The shaded area is covered if
the $ \lambda^{\prime}$ couplings are varied over
their entire allowed range. }
\label{f:lpno}
\end{figure}

If neutrinos have only Standard Model interactions 
then the expected number of muon events is fixed\footnote{Recall
we assume that, but for $\theta_{13}$ and the mass hierarchy, the
other neutrino mass and mixing parameters are known.} for  a
particular value of $\theta_{13}$ with either normal or inverted
hierarchy as may be seen from the solid lines in Fig.
\ref{f:lpno}. The vast difference for the alternate hierarchies
picks out such long baseline experiments as good laboratories for
addressing this open question of the neutrino mass spectrum.

If non-standard interactions are present then, depending on their
coupling strength, the picture can change dramatically.   In Fig.
\ref{f:lpno}, the shaded region corresponds to the allowed values
when SUSY FCNC and FDNC interactions are at play. It is obtained
by letting the $\lambda^{\prime}$ couplings\footnote{In fact, we have
chosen the subscript $m$ in the $\lambda^{\prime}$ couplings in eq.
(\ref{e:R1}) to be any one of 1,2, or 3.} vary over their entire
allowed range   -- both positive and negative -- given in eq.
(\ref{e:lim}), subject to the further constraints on particular
products. 

It is seen that to a significant extent the distinguishability of
the two hierarchies is obstructed by the $\Rsl$ interactions
unless the number of events is more than about 60.  Also, the
one-to-one correspondence is lost between $\theta_{13}$ and the
number of events and, at best, a lower bound can now be placed on
$\theta_{13}$ from the observed number. Of course, if the
neutrino mass hierarchy is known from other experiments, then
this lower bound can be strengthened, especially for the inverted
hierarchy.

It is also noteworthy  that for some values of
$\lambda^{\prime}$-couplings there may be more events than can be
expected from the Standard Model interactions, no matter what the
value of $\theta_{13}$. Thus, observation of more than 161 (5)
events for the normal (inverted) hierarchy would be a  clear
signal of new physics.

\subsection{Constraining $\lambda^{\prime}$ }

If $\theta_{13}$ is determined from other experiments then it
will be easier to look for  non-standard signals from this
$\beta$ beam experiment.  However, even if the precise value
remains unknown at the time, considering the upper bound on
$\theta_{13}$ one may tighten the constraints on the
$\lambda^{\prime}$ couplings.  Fig. \ref{f:lpno} reflects the
overall sensitivity of the event rate to the $\Rsl$ interactions
obtained by letting all RPVSM couplings vary over their entire
allowed ranges. In this subsection, we want to be more specific
and ask how the event rate depends on any chosen
$\lambda^{\prime}$ coupling.

At the outset, it may be worth recalling that the BELLE bound on
$\tau \rightarrow \mu \pi^0$ \cite{belltau} severely limits the
products $\lambda^{\prime}_{2m1} \lambda^{\prime}_{3m1}$ and
$\lambda^{\prime}_{21m} \lambda^{\prime}_{31m}$. Thus,  $R_{23}$
can be dropped in the effective neutrino mass matrix eq.
(\ref{e:m2}).  If only $\lambda^{\prime}_{2m1}$ and/or
$\lambda^{\prime}_{21m}$ ($\lambda^{\prime}_{3m1}$ and/or
$\lambda^{\prime}_{31m}$) is non-zero, then $R_{22}$ ($R_{33}$)
alone receives an RPVSM contribution.  Both $R_{22}$ and $R_{33}$
can be simultaneously non-zero if $\lambda^{\prime}_{21m}$ and
$\lambda^{\prime}_{3m1}$ (or $\lambda^{\prime}_{2m1}$ and
$\lambda^{\prime}_{31m}$) are non-zero at the same time.

\begin{figure}[thb]
\hskip -0.4cm 
\vskip 1.85cm
\psfig{figure=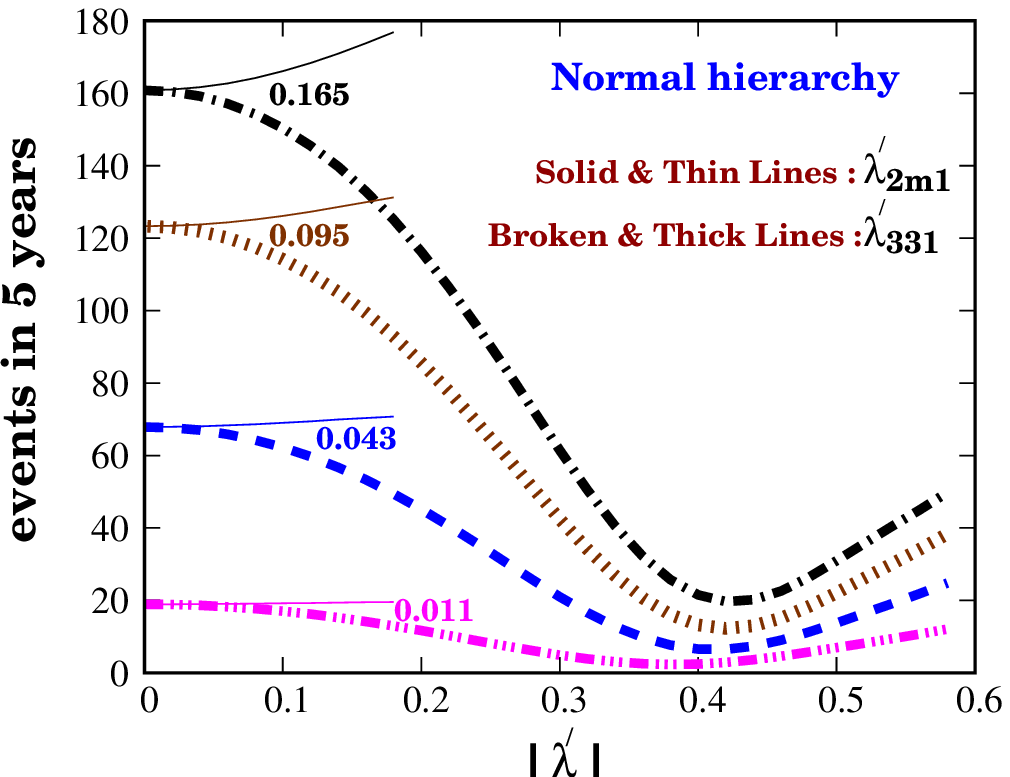,width=8.0cm,height=7.0cm,angle=0}
\vskip -7.0cm
\hskip 8.63cm
\psfig{figure=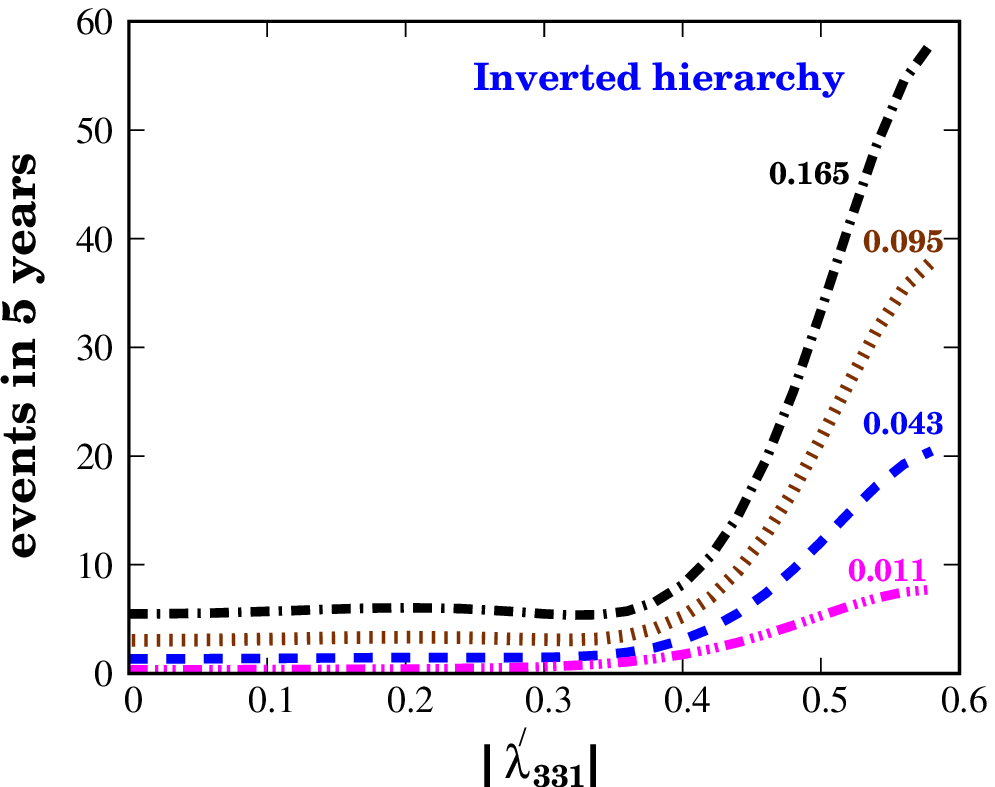,width=8.0cm,height=7.0cm,angle=0}
\caption{\sf \small
The number of events as a function of a coupling $|\lambda^{\prime}|$,
present singly, for the 
 normal (left panel) and inverted (right panel) 
hierarchies. The thick (thin) lines are for
$|\lambda^{\prime}_{331}|$ ($|\lambda^{\prime}_{2m1}|$,
$m=$2,3).The chosen  $\sin^2 2 \theta_{13}$ are indicated next to
the curves.}
\label{f:bndl3}
\end{figure}

In the light of this, we consider the situation where only of the
above $\Rsl$ coupling is non-zero.  In such an event, only one of
$R_{22}$, $R_{33}$ is non-zero. The dependence of the number of
events  on a non-zero $\lambda^{\prime}_{331}$ or
$\lambda^{\prime}_{2m1}$, for a chosen $\sin^2 2\theta_{13}$, can
be seen from Fig.
\ref{f:bndl3}. In this figure, we use the fact that if only one
of these $\Rsl$ couplings is non-zero, it appears in the results
through $|\lambda^{\prime}|$. For the normal hierarchy, the
curves for $\lambda^{\prime}_{2m1}$, for $m =$ 2,3, are
terminated at the maximum allowed value of 0.18. Fig.
\ref{f:bndl3} can also be used for
$\lambda^{\prime}_{321}$,
$\lambda^{\prime}_{31m}$ and $\lambda^{\prime}_{21m}$, bearing in mind
their different upper bounds.
For the inverted hierarchy, the number of events is 
small for $\lambda^{\prime}_{2m1}$ and
$\lambda^{\prime}_{21m}$ and insensitive to the magnitude of the
coupling. These are not shown. It is seen that for the  
normal hierarchy there is a good chance to determine the $\Rsl$
couplings from the number of events. In fact, if the number of
events is less than about 50 there is a disallowed region for
$|\lambda^{\prime}|$, while for larger numbers there is only an upper
bound. For the inverted hierarchy, more than about five events
will set a lower bound on the coupling.


\subsection{Effect of $\lambda$} 
\begin{figure}[hbt]
\hskip -0.4cm
\vskip -0.00cm
\psfig{figure=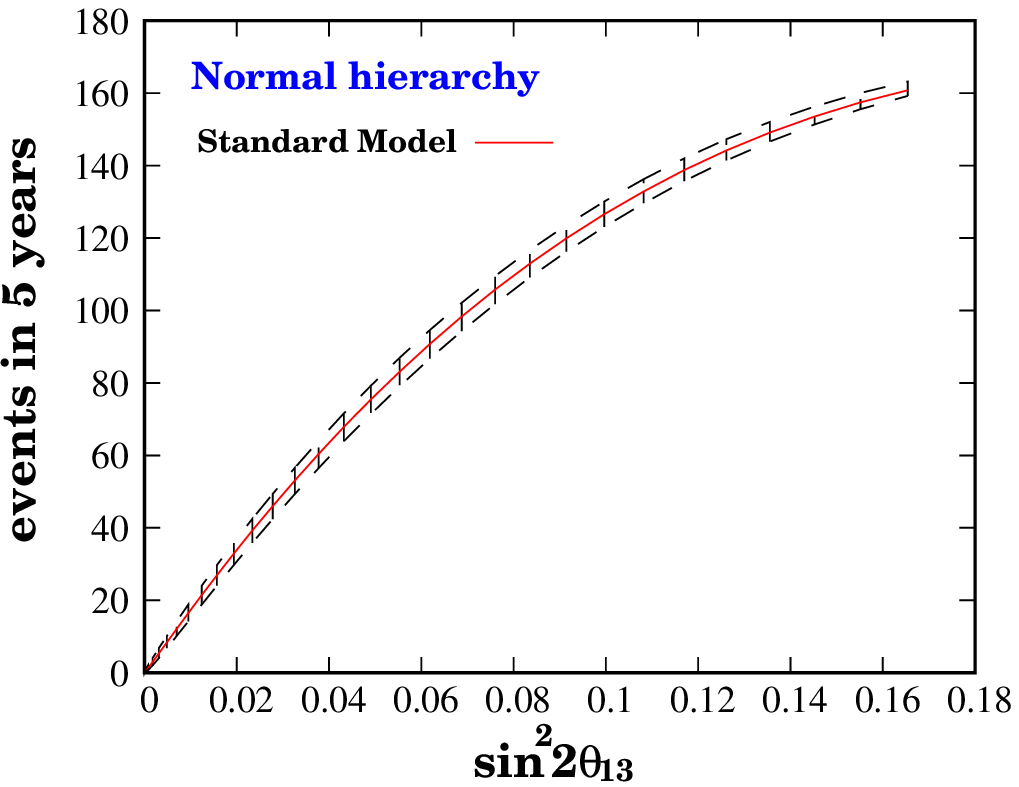,width=8.0cm,height=7.0cm,angle=0}
\vskip -7.00cm
\hskip 8.63cm
\psfig{figure=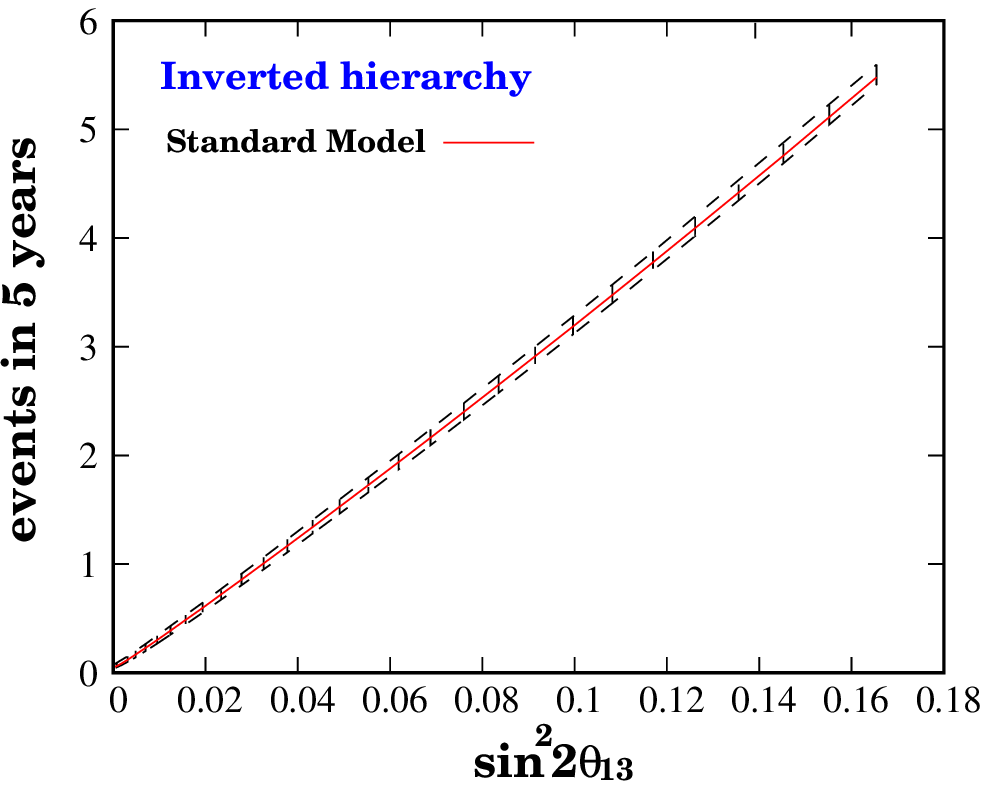,width=8.0cm,height=7.0cm,angle=0}
\caption{\sf \small
Number of muon events for normal and inverted mass hierarchies as
a function of $\sin^2 2\theta_{13}$ for a five-year ICAL run.
The solid lines correspond to the absence of any non-Standard 
neutrino interaction. The shaded area is covered if
the $ \lambda$ couplings are varied over
their entire allowed range. }
\label{f:lno}
\end{figure}

The $\lambda$ couplings which can
contribute in eq. (\ref{e:rl}) have strong existing bounds
\cite{rparity} and their contribution to $R$ is rather small  in
comparison to $\sqrt{2} G_F n_e$.  Among them, the bounds
$\lambda_{121} < 0.05 $ and $\lambda_{321} < 0.07 $ for
$m_{\tilde l} = 100$ GeV are relatively less stringent
\cite{rparity}. We show their very modest impact in Fig.
\ref{f:lno}. It is clear from this figure that (a) the
$\lambda$-type couplings cannot seriously deter the extraction of
$\theta_{13}$ or the determination of the neutrino mass
hierarchy, and (b) when $\theta_{13}$ is known in future it will
still not be possible to constrain these couplings through long
baseline experiments.

\section{Conclusions}

R-parity violating supersymmetry is among several
extensions of the Standard Model crying out for experimental
verification. The model has flavour diagonal and flavour changing
neutral currents which can affect neutrino masses and mixing and
can leave their imprints in long baseline experiments. This is
the focus of this work.

We consider a $\beta$-beam experiment with the source at CERN and
the detector at INO. We find that the $\Rsl$ interactions may
obstruct a clean extraction of the mixing angle $\theta_{13}$ or
determination of the mass hierarchy unless the bounds on the
$\lambda^{\prime}$ couplings are tightened. On the other hand, one
might be able to see a clean signal of new physics. Here, the
long baseline comes as a boon over experiments like MINOS which
cover shorter distances. Two experiments of these contrasting
types, taken together, can expose the presence of a non-standard
interaction like RPVSM.

There are other non-standard models \cite{david} where
four-fermion neutrino couplings with greater strength
have been invoked. The signals we consider
will be much enhanced in such cases.

Our results are presented for the $CP$ conserving case. As
$\theta_{13}$ is small, the $CP$ violating effect is expected to
be suppressed. We have checked this for the Standard Model,
where the `magic' nature of the baseline \cite{magic} also plays
a role.

Finally, in this paper we have restricted ourselves to a
$\beta$-beam neutrino source. Much the same could be done for
antineutrinos as well; then the signs of all terms in $R$ -- see
eq. (\ref{e:m2}) --  will be reversed. It follows from eq. (\ref{e:mmax})
that $\theta_{13}^m$ can then be maximal only for the
inverted hierarchy and as such more events are expected here than
in the normal hierarchy. Broadly, results similar to the ones
presented here with neutrinos can be obtained with antineutrinos
if normal hierarchy is replaced by inverted hierarchy and
{\em vice-versa}.

\vskip 1.0cm

\noindent
{\large{\bf {Acknowledgments}}}\\ 

R. Adhikari acknowledges hospitality of the Harish-Chandra
Research Institute under the DAE X-th plan project on collider
physics while this work was done. \\


\end{document}